\documentstyle[12pt]{article}

\def\hybrid{\topmargin -20pt    \oddsidemargin 0pt
        \headheight 0pt \headsep 0pt
        \textwidth 6.25in       
        \textheight 9.5in       
        \marginparwidth .875in
        \parskip 5pt plus 1pt   \jot = 1.5ex}
\def\cQ{{\cal Q}}
\def\cG{{\cal G}}
\def\cL{{\cal L}}

\def\cH{{\cal H}}
\def\ket#1{|{#1}\rangle}
\def\noi{\noindent}
\def\half{{1\over2}}
\def\baselinestretch{1.2}

\catcode`\@=11

\def\marginnote#1{}
\def\draftlabel#1{{\@bsphack\if@filesw {\let\thepage\relax
   \xdef\@gtempa{\write\@auxout{\string
      \newlabel{#1}{{\@currentlabel}{\thepage}}}}}\@gtempa
   \if@nobreak \ifvmode\nobreak\fi\fi\fi\@esphack}
        \gdef\@eqnlabel{#1}}
\def\@eqnlabel{}
\def\@vacuum{}
\def\draftmarginnote#1{\marginpar{\raggedright\scriptsize\tt#1}}

\def\draft{\oddsidemargin -.2truein
        \def\@oddfoot{\sl preliminary draft \hfil
        \rm\thepage\hfil\sl\today\quad\militarytime}
        \let\@evenfoot\@oddfoot \overfullrule 3pt
        \let\label=\draftlabel
        \let\marginnote=\draftmarginnote
   \def\@eqnnum{(\theequation)\rlap{\kern\marginparsep\tt\@eqnlabel}%
\global\let\@eqnlabel\@vacuum}  }

\def\preprint{\twocolumn\sloppy\flushbottom\parindent 2em
        \leftmargini 2em\leftmarginv .5em\leftmarginvi .5em
        \oddsidemargin -.5in    \evensidemargin -.5in
        \columnsep .4in \footheight 0pt
        \textwidth 10.in        \topmargin  -.4in
        \headheight 12pt \topskip .4in
        \textheight 6.9in \footskip 0pt
        \def\@oddhead{\thepage\hfil\addtocounter{page}{1}\thepage}
        \let\@evenhead\@oddhead \def\@oddfoot{} \def\@evenfoot{} }

\def\numberbysection{\@addtoreset{equation}{section}
        \def\theequation{\thesection.\arabic{equation}}}

\def\underline#1{\relax\ifmmode\@@underline#1\else
        $\@@underline{\hbox{#1}}$\relax\fi}

\def\titlepage{\@restonecolfalse\if@twocolumn\@restonecoltrue
\onecolumn
     \else \newpage \fi \thispagestyle{empty}\c@page\z@
        \def\thefootnote{\fnsymbol{footnote}} }

\def\endtitlepage{\if@restonecol\twocolumn \else \newpage \fi
        \def\thefootnote{\arabic{footnote}}
        \setcounter{footnote}{0}}  

\catcode`@=12
\relax

\def\figcap{\section*{Figure Captions\markboth
        {FIGURECAPTIONS}{FIGURECAPTIONS}}\list
        {Figure \arabic{enumi}:\hfill}{\settowidth\labelwidth{Figure
999:}
        \leftmargin\labelwidth
        \advance\leftmargin\labelsep\usecounter{enumi}}}
 \relax
\def\tablecap{\section*{Table Captions\markboth
        {TABLECAPTIONS}{TABLECAPTIONS}}\list
        {Table \arabic{enumi}:\hfill}{\settowidth\labelwidth{Table
999:}
        \leftmargin\labelwidth
        \advance\leftmargin\labelsep\usecounter{enumi}}}
 \relax
\def\reflist{\section*{References\markboth
        {REFLIST}{REFLIST}}\list
        {[\arabic{enumi}]\hfill}{\settowidth\labelwidth{[999]}
        \leftmargin\labelwidth
        \advance\leftmargin\labelsep\usecounter{enumi}}}
 \relax
\makeatletter
\newcounter{pubctr}
\def\publist{\@ifnextchar[{\@publist}{\@@publist}}
\def\@publist[#1]{\list
        {[\arabic{pubctr}]\hfill}{\settowidth\labelwidth{[999]}
        \leftmargin\labelwidth
        \advance\leftmargin\labelsep
        \@nmbrlisttrue\def\@listctr{pubctr}
        \setcounter{pubctr}{#1}\addtocounter{pubctr}{-1}}}
\def\@@publist{\list
        {[\arabic{pubctr}]\hfill}{\settowidth\labelwidth{[999]}
        \leftmargin\labelwidth
        \advance\leftmargin\labelsep
        \@nmbrlisttrue\def\@listctr{pubctr}}}
 \relax
\makeatother
\newskip\humongous \humongous=0pt plus 1000pt minus 1000pt

\newif\ifdtup

\font\Scbig=cmss10 scaled\magstep1
\font\Scscr=cmss8 scaled\magstep1
\font\Scscrscr=cmss8
\newfam\Scfam
\textfont\Scfam=\Scbig
\scriptfont\Scfam=\Scscr
\scriptscriptfont\Scfam=\Scscrscr
\def\Sc{\fam\Scfam}

\relax
\hybrid
\def\lvm{\leavevmode\hbox to\parindent{\hfill}}

\def\thefootnote{\fnsymbol{footnote}}
\def\BE{\begin{equation}}
\def\EE{\end{equation}}
\def\BA{\begin{eqnarray}}
\def\EA{\end{eqnarray}}

\def\D{\Delta}

\def\tt{\bar\tau}

\def\lvm{\leavevmode\hbox to\parindent{\hfill}}
\def\bar{\overline}
\def\req#1{(\ref{#1})}

\def\L{\left}
\def\R{\right}

\def\BE{\begin{equation}}
\def\EE{\end{equation} \vskip 0.30\baselineskip}
\def\BA{\begin{array}}
\def\EA{\end{array}}

\def\noi{\noindent}

\def\frac#1#2{{\textstyle{{#1}\over{#2}}}}
\def\half{{1\over2}}

\def\Kr#1{\delta_{{#1},0}}
\def\ket#1{|{#1}\rangle}

\def\cA{{\cal A}}
\def\cG{{\cal G}}
\def\cH{{\cal H}}

\def\cL{{\cal L}}

\def\cQ{{\cal Q}}

\def\open#1{\mbox{{\bf{#1}}}}

\def\oZ{{\open Z}}

\def\ctop{{\Sc c}}
\def\htop{{\Sc h}}

\def\svec{singular vector}

\def\kp{\ket\phi}

\def\ie{{\it i.e.}}

\def\Qz{\cQ_0}
\def\Gz{\cG_0}
\def\Qn{$\Qz$}
\def\Gn{$\Gz$}
\def\kc{{\ket{\chi}}}
\def\kcc#1#2#3{{\kc_{#1}^{({#2}){#3}}}}

\newif\ifold \oldtrue \def\new{\oldfalse}
\let\ssection=\section
\def\section{\setcounter{equation}{0}\ssection}

\begin{document}
\renewcommand{\theequation}{\thesection.\arabic{equation}}
\newcommand{\beq}{\begin{equation}}
\newcommand{\eeq}[1]{\label{#1}\end{equation}}
\newcommand{\ber}{\begin{eqnarray}}
\newcommand{\eer}[1]{\label{#1}\end{eqnarray}}
\begin{titlepage}
\begin{center}

\hfill IMAFF-FM-98/05,  NIKHEF-98-005\\
\hfill hep-th/9802204
\vskip .4in

{\large \bf Construction Formulae for Singular Vectors of the Topological 
N=2 Superconformal Algebra}
\vskip .4in

{\bf Beatriz Gato-Rivera}
\vskip .3in

 {\em Instituto de Matem\'aticas y F\'\i sica Fundamental, CSIC,\\
 Serrano 123, Madrid 28006, Spain} \footnote{e-mail address:
bgato@pinar1.csic.es}\\
\vskip .2in

{\em NIKHEF-H, Kruislaan 409, NL-1098 SJ Amsterdam, The Netherlands}\\

\vskip 1.0in

\end{center}

\begin{center} {\bf ABSTRACT } \end{center}
\begin{quotation}
The Topological N=2 Superconformal algebra has 29 different types of singular 
vectors (in complete Verma modules) distinguished by the relative U(1) charge 
and the BRST-invariance properties of the vector and of the primary on which 
it is built. Whereas one of these types only exists at level zero, the 
remaining 28 types exist for general levels and can be constructed already at 
level 1. In this paper we write down one-to-one mappings between 16 of these 
types of topological singular vectors and the singular vectors of the 
Antiperiodic NS algebra. As a result one obtains construction formulae for 
these 16 types of topological singular vectors using the construction 
formulae for the NS singular vectors due to D\"orrzapf.

\end{quotation}
\vskip 1.5cm

February 1998
\end{titlepage}

\def\baselinestretch{1.2}
\baselineskip 17 pt
\section{Introduction and Notation}\lvm

The N=2 Superconformal algebras provide the symmetries underlying the
N=2 strings \cite{Ade}\cite{Marcus}. These seem to be related to M-theory 
since many of the basic objects of M-theory are realized in the 
heterotic (2,1) N=2 strings \cite{Marti}. In addition, the topological
version of the algebra is realized in the world-sheet of the bosonic
string \cite{BeSe}, as well as in the world-sheet of the superstrings
\cite{BLNW}.

Recently it has been shown \cite{BJI6} that the singular vectors of the
Topological N=2 algebra can be classified in 29 types,
in complete Verma modules, taking into account the relative U(1) charge
and the BRST-invariance properties of the vector itself and of the primary
on which it is built. In ref. \cite{BJI6} the whole set of singular vectors
was explicitely constructed at level 1 (28 types since one type exists only
at level 0), whereas the rigorous proofs that these types are the only 
possible ones will be given in \cite{DB2}. 

In this paper we intend to bring to the reader's attention the fact that 
one can write down 
one-to-one mappings between 16 of these types of topological 
singular vectors and the singular vectors of the NS algebra. 
As a bonus one obtains construction formulae for the 16 types of topological 
singular vectors using the construction formulae for the NS singular vectors
due to D\"orrzapf \cite{Doerr1}\cite{Doerr2}. In section 2 we discuss 
the basic ingredients to derive the mappings between the 
topological singular vectors and the NS singular vectors. In section 3
we write down these mappings, which turn into construction formulae for 
the topological singular vectors once the NS singular vectors are expressed 
in terms of their construction formulae themselves. Some final remarks
are made in section 4.

\vskip .35in
\noi
{\bf Notation}
\vskip .17in
\noi
{\it Highest weight (h.w.) states} denote states annihilated by all
the positive modes of the generators of the algebra, \ie\
${\ } \cL_{n \geq 1} \kc =  \cH_{n \geq 1} \kc =  {\cG}_{n \geq 1} \kc
=  {\cQ}_{n \geq 1} \kc = 0 {\ }$.

\noi
{\it Primary states} denote non-singular h.w. states. 

\noi
{\it Secondary or descendant states} denote states obtained by acting on 
the h.w. states with the negative modes of the generators of the algebra
and with the fermionic zero modes \Qn\ and \Gn\ . The fermionic zero 
modes can also interpolate between two h.w. states at the same footing
(two primary states or two singular vectors).

\noi
{\it Chiral topological states} $\kc^{G,Q}$ are states
annihilated by both $\cG_0$ and $\cQ_0$. 

\noi
{\it $\cG_0$-closed topological states} $\kc^G$ denote  
non-chiral states annihilated by $\cG_0$. 

\noi
{\it $\cQ_0$-closed topological states} $\kc^Q$ denote  
non-chiral states annihilated by $\cQ_0$
(they are BRST-invariant since $\cQ_0$ is the BRST charge).

\noi
{\it No-label topological states} $\kc$ denote states  
that cannot be expressed as linear combinations of $\cG_0$-closed
and $\cQ_0$-closed states.

\noi
{\it The Verma module} associated to a h.w. state consists of the
h.w. state plus the set of secondary states built on it. For some
Verma modules the h.w. state is degenerate, the fermionic zero
modes interpolating between the two h.w. states.

\noi
{\it Singular vectors} are h.w. zero-norm states.

\noi
{\it Secondary singular vectors} are singular vectors built on singular
vectors. The level-zero secondary singular vectors cannot ``come back" to 
the singular vectors on which they are built by acting with \Gn\ or \Qn\ .

\noi
The Topological N=2 superconformal algebra will be denoted as 
{\it the Topological algebra}.

\noi
The Antiperiodic N=2 superconformal algebra will be denoted as 
{\it the NS algebra}.

\section{Basic Concepts}\lvm

\noi
{\bf The Topological algebra and the topological twists}

\vskip .13in

The algebra obtained by applying the topological twists on the
NS algebra reads \cite{DVV}

\BE\new\BA{lclclcl}
\L[\cL_m,\cL_n\R]&=&(m-n)\cL_{m+n}\,,&\qquad&[\cH_m,\cH_n]&=
&{\ctop\over3}m\Kr{m+n}\,,\\
\L[\cL_m,\cG_n\R]&=&(m-n)\cG_{m+n}\,,&\qquad&[\cH_m,\cG_n]&=&\cG_{m+n}\,,
\\
\L[\cL_m,\cQ_n\R]&=&-n\cQ_{m+n}\,,&\qquad&[\cH_m,\cQ_n]&=&-\cQ_{m+n}\,,\\
\L[\cL_m,\cH_n\R]&=&\multicolumn{5}{l}{-n\cH_{m+n}+{\ctop\over6}(m^2+m)
\Kr{m+n}\,,}\\
\L\{\cG_m,\cQ_n\R\}&=&\multicolumn{5}{l}{2\cL_{m+n}-2n\cH_{m+n}+
{\ctop\over3}(m^2+m)\Kr{m+n}\,,}\EA\qquad m,~n\in\oZ\,.\label{topalgebra}
\EE

\noi
where $\cL_m$ and $\cH_m$ are the bosonic generators corresponding
to the energy momentum tensor (Virasoro generators)
 and the topological $U(1)$ current respectively, while
$\cQ_m$ and $\cG_m$ are the fermionic generators corresponding
to the BRST current and the spin-2 fermionic current
respectively. The eigenvalues of $\cL_0$ and $\cH_0$ correspond to
the conformal weight $\D$ and the U(1) charge $\htop$ of the states.
The ``topological" central charge $\ctop$ is the central charge 
corresponding to the NS algebra. This algebra is 
topological because the Virasoro generators can be expressed as
$\cL_m={1\over2}\{\cG_m,\cQ_0\}$, where $\cQ_0$ is the BRST charge. This
implies, as is well known, that the correlators of the fields do not
depend on the metric.

\vskip .15in

The two possible topological twists of the NS 
superconformal generators are:

\BE\new\BA{rclcrcl}
\cL^{(\pm)}_m&=&\multicolumn{5}{l}{L_m \pm \half(m+1)H_m\,,}\\
\cH^{(\pm)}_m&=&\pm H_m\,,&{}&{}&{}&{}\\
\cG^{(\pm)}_m&=&G_{m+\half}^{\pm}\,,&\qquad &\cQ_m^{(\pm)}&=
&G^{\mp}_{m-\half} \,,\label{twa}\EA\EE

\noi
These twists, which we denote as $T_W^{\pm }$,  
are mirrored under the interchange $H_m \leftrightarrow -H_m$, 
${\ } G^{+}_r \leftrightarrow G^{-}_r$. Observe that the h.w. conditions
$G^{\pm}_{1/2}\, \ket{\chi_{NS}} = 0$ of the NS algebra 
read $\Gz \kc = 0$ after the corresponding twists.
Therefore, any h.w. state of the NS algebra results
in a \Gn-closed or chiral state of the Topological algebra, which is also
h.w. as the reader can easily verify by inspecting the twists
\req{twa}. Conversely, any \Gn-closed or chiral h.w.
topological state (and only these) transforms into a h.w. state
of the NS algebra.

\vskip .17in
\noi
{\bf Topological states and topological singular vectors}

\vskip .13in

{ }From the anticommutator $\{ \cQ_0, \cG_0\} = 2 \cL_0 $ one deduces 
\cite{BJI6} that a topological state (primary or secondary)
 with non-zero conformal weight can be either
 \Gn-closed, or \Qn-closed, or a linear combination of both types. 
The topological states with zero conformal weight, however, can be 
\Qn-closed, or \Gn-closed, or chiral, or no-label.

As a first classification of the topological secondary states one considers
their level $l$, their {\it relative} U(1) charge $q$
and their transformation properties under
\Qn\ and \Gn\ (BRST-invariance properties). The level $l$ 
and the relative charge $q$ are defined as the difference between
the conformal weight and U(1)
charge of the secondary state and the conformal weight and U(1) charge
of the primary state on which it is built.
Hence the topological secondary states will be denoted as $\kc_l^{(q)G}$
($\cG_0$-closed), $\kc_l^{(q)Q}$ ($\cQ_0$-closed), 
$\kc_l^{(q)G,Q}$ (chiral), and $\kc_l^{(q)}$ (no-label).
For convenience we will also indicate the conformal weight $\D$,
the U(1) charge $\htop$, and the BRST-invariance properties
of the primary state on which the secondary is built. Observe that
the conformal weight and the total U(1) charge of the secondary
states are given by $\D + l$ and $\htop+q$, respectively. 

\vskip .15in

In complete Verma modules there are 29 types of topological singular vectors 
\cite{BJI6}, a given type being defined by the relative charge $q$, the
BRST-invariance properties of the state itself, and the
BRST-invariance properties of the primary on which it is built.
The singular vectors with non-zero conformal weight, 
$\D+l \neq 0$, are linear combinations
of \Gn-closed and \Qn-closed singular vectors \cite{BJI6}, therefore 
we can restrict ourselves to these types with well defined BRST-invariance
properties. The singular vectors with zero conformal weight, 
$\D+l=0$, can also be chiral or no-label.

There are three different types of topological primaries giving rise
to complete Verma modules: \Gn-closed primaries $\ket{\D,\htop}^G$, 
\Qn-closed primaries $\ket{\D,\htop}^Q$, and no-label primaries
$\ket{0,\htop}$. We do not consider primaries which are linear
combinations of two or more of these types, neither primaries of these
types with additional constraints (like chiral primaries) 
giving rise to incomplete Verma modules.
The topological \svec s in no-label Verma modules 
(9 types) cannot be mapped to NS \svec s.
The remaining 20 types are distributed in the following way:

- Ten types built on $\cG_0$-closed primaries
$\ket{\D,\htop}^G$:

\BE
\begin{tabular}{r|l l l l}
{\ }& $q=-2$ & $q=-1$ & $q=0$ & $q=1$\\
\hline\\
\Gn-closed & $-$ & $\kc_l^{(-1)G}$ & $\kc_l^{(0)G}$ & $\kc_l^{(1)G}$\\
\Qn-closed & $\kc_l^{(-2)Q}$ & $\kc_l^{(-1)Q}$ & $\kc_l^{(0)Q}$ & $-$ \\
chiral & $-$ & $\kc_l^{(-1)G,Q}$ & 
$\kc_l^{(0)G,Q}$ & $-$ \\
no-label & $-$ & $\kc_l^{(-1)}$ &
$\kc_l^{(0)}$ & $-$\\
\end{tabular}
\label{tabl2}
\EE

- Ten types built on $\cQ_0$-closed primaries
$\ket{\D,\htop}^Q$:

\BE
\begin{tabular}{r|l l l l}
{\ }& $q=-1$ & $q=0$ & $q=1$ & $q=2$\\
\hline\\
$\cG_0$-closed & $-$ & $\kc_l^{(0)G}$ & $\kc_l^{(1)G}$ & $\kc_l^{(2)G}$\\
$\cQ_0$-closed & $\kc_l^{(-1)Q}$ & $\kc_l^{(0)Q}$ & $\kc_l^{(1)Q}$ & $-$ \\
chiral & $-$ & $\kc_l^{(0)G,Q}$ & $\kc_l^{(1)G,Q}$ & 
$-$ \\
no-label & $-$ & $\kc_l^{(0)}$ &
$\kc_l^{(1)}$ & $-$\\
\end{tabular}
\label{tabl3}
\EE

\vskip .17in

For $\D \neq 0$ the h.w. vector of the Verma module is degenerate: there 
is one \Gn-closed primary state as well as one \Qn-closed primary state,
\Qn\ and \Gn\ interpolating between them.
As a result, for $\D \neq 0$ the singular vectors of table \req{tabl2}
are equivalent to singular vectors of table \req{tabl3} with a shift
on the U(1) charges (for the details see ref. \cite{BJI6}). In particular,
the charged (uncharged) chiral \svec s of table \req{tabl2} are equivalent
to the uncharged (charged) chiral \svec s of table \req{tabl3}. With 
the exception of the no-label \svec s, all other 16 types of topological 
\svec s in tables \req{tabl2} and \req{tabl3} can be mapped to NS \svec s,
as we will see in next section.

An important observation is that chiral singular vectors
$\kc_{l}^{(q)G,Q}\,$ can be regarded as particular cases of \Gn-closed
singular vectors $\kc_{l}^{(q)G}\,$ and  also as particular cases
of \Qn-closed singular vectors $\kc_{l}^{(q)Q}\,$. That is, 
some \Gn-closed and \Qn-closed singular vectors may ``become" chiral 
(although not necessarily, depending on the case) when the conformal weight 
of the singular vector turns out to be zero, \ie\ $\D+l=0$.

\vskip .17in
\noi
{\bf The fermionic zero modes}

\vskip .13in

Most topological singular vectors come in pairs at the same level in
the same Verma module, differing by one unit of relative charge. The
reason is that the fermionic zero modes \Gn\ and \Qn\ acting on a singular
vector produce another singular vector \cite{BJI3}\cite{BJI6}, 
as can be checked straightforwardly using the Topological algebra 
\req{topalgebra}. Therefore only chiral singular vectors can be ``alone",
whereas the no-label singular vectors are accompanied by three, rather 
than one, singular vectors at the same level in the same Verma module.
To be precise, inside a given Verma module $V(\Delta,\htop)$ and for 
a given level $l$ the topological singular vectors 
with non-zero conformal weight are connected
by the action of \Qn\ and \Gn\ as:
\begin{eqnarray}  \Qz \, \kcc{l}{q}{G} \to \kcc{l}{q-1}{Q} , 
 \qquad\Gz \,\kcc{l}{q}{Q} \to \kcc{l}{q+1}{G}\, , \label{GQh}
\end{eqnarray}

\noi
where the arrows can be reversed (up to constants), using \Gn\ and
\Qn\ respectively, since the conformal weight of the singular vectors
is different from zero, \ie\ $\D+l\neq0$. Otherwise, on the right-hand side
of the arrows one obtains {\it chiral secondary} singular vectors
which cannot ``come back" to
the singular vectors on the left-hand side:  
\begin{eqnarray}  \Qz \,\kcc{l=-\D}{q}{G} \to \kcc{l=-\D}{q-1}{G,Q} , 
 \qquad\Gz \,\kcc{l=-\D}{q}{Q} \to \kcc{l=-\D}{q+1}{G,Q}\, . \label{GQch}
\end{eqnarray}

Regarding no-label singular vectors 
$\kc_l^{(q)}$, they always satisfy $\D+l=0$. The action of \Gn\ and \Qn\
on a no-label singular vector produce three singular vectors:
 \begin{eqnarray} \Qz \,\kcc{l=-\D}{q}{ } \to \kcc{l=-\D}{q-1}{Q} ,  
 \quad\Gz \,\kcc{l=-\D}{q}{ } \to \kcc{l=-\D}{q+1}{G} , \quad 
 \Gz \, \Qz \,\kcc{l=-\D}{q}{ } \to \kcc{l=-\D}{q}{G,Q} \,. \label {QGnh} 
 \end{eqnarray}

\noi
All three are secondary \svec s which cannot come back to the no-label
\svec $\kcc{l=-\D}{q}{ }$ by acting with \Gn\ and \Qn\ .

Hence \Gn\ and \Qn\ interpolate between two singular vectors
with non-zero conformal weight, in both directions, whereas they 
produce secondary singular vectors when acting on singular vectors
with zero conformal weight.

\vskip 0.17in
\noi 
{\bf The universal odd spectral flow} $\cA$

\vskip .13in

The universal odd spectral flow automorphism $\cA_1$, 
denoted simply as $\cA$, transforms all kinds of
primary states and singular vectors back into primary states and
singular vectors, mapping chiral states to chiral states.
It is given by \cite{BJI3} \cite{B1}

\BE\new\BA{rclcrcl}
\cA \, \cL_m \, \cA^{-1}&=& \cL_m - m\cH_m\,,\\
\cA \, \cH_m \, \cA^{-1}&=&-\cH_m - {\ctop\over3} \delta_{m,0}\,,\\
\cA \, \cQ_m \, \cA^{-1}&=&\cG_m\,,\\
\cA {\ } \cG_m \, \cA^{-1}&=&\cQ_m\,.\
\label{autom} \EA\EE

\noi
with  $\cA^{-1} = \cA$. 
It transforms the $(\cL_0,\cH_0)$ eigenvalues $(\D,\htop)$
of the states as $(\D,-\htop-{\ctop\over3})$, 
reversing the relative charge of the secondary states and 
leaving the level 
invariant, as a consequence. In addition, $\cA$ also reverses 
the BRST-invariance properties of the states (primary as well as secondary) 
mapping \Gn-closed (\Qn-closed) states into \Qn-closed (\Gn-closed)
states, and chiral states into chiral states. Hence the action of $\cA$ 
results in the following mappings between singular vectors in
different Verma modules:
\begin{eqnarray}
{\cal A}\, \kcc{l,\, \ket{\D,\,\htop}^G}{q}{G} \to \kcc{l,\,
\ket{\D,-\htop-{\ctop\over3}}^Q}{-q}{Q}\ , \ \ \
\ \ {\cal A} \,\kcc{l,\, \ket{\D,\,\htop}^G}{q}{Q} \to \kcc{l,
\, \ket{\D,-\htop-{\ctop\over3}}^Q}{-q}{G} ,\nonumber \\
{\cal A}\, \kcc{l,\, \ket{-l,\,\htop}^G}{q}{G,Q} \to \kcc{l,\,
\ket{-l,-\htop-{\ctop\over3}}^Q}{-q}{G,Q}\ , \ \ 
\ \ {\cal A} \,\kcc{l,\, \ket{-l,\,\htop}^G}{q}{ } \to \kcc{l,
\, \ket{-l,-\htop-{\ctop\over3}}^Q}{-q}{ },\nonumber\\ 
{\cal A}\, \kcc{l,\, \ket{0,\,\htop}}{q}{G} \to \kcc{l,\,
\ket{0,-\htop-{\ctop\over3}}}{-q}{Q}\ , \ \ \ \ \ \ \ \
\ \ {\cal A}\, \kcc{l,\, \ket{0,\,\htop}}{q}{Q} \to \kcc{l,\,
\ket{0,-\htop-{\ctop\over3}}}{-q}{G} ,
\label{AADh}
\end{eqnarray}

\noi
and their inverses. 

\section{Construction Formulae}\lvm

About four years ago construction formulae for the singular vectors of 
the NS algebra were computed by D\"orrzapf. Using the ``fusion method" 
\cite{fusion} explicit formulae were obtained 
\cite{Doerr1} for all the charged singular
vectors (which only exist for $q=\pm1$), and for 
a class of uncharged singular vectors analogous to the BSA singular
vectors of the Virasoro algebra \cite{BSA}. Later using the ``analytic
continuation method" \cite{analyt} explicit formulae were obtained
for all the uncharged singular vectors \cite{Doerr2}.

In what follows we will show that 
these construction formulae for the NS singular vectors also provide
construction formulae for 16 types of topological singular vectors 
since one can write down mappings from the NS singular vectors to these 16 
types of topological singular vectors; one-to-one mappings in particular.
We will proceed in the following way. First we will construct
the ``box" diagrams obtained by the actions of
the fermionic zero modes \Gn, \Qn\ and the action of the spectral flow 
automorphism $\cA$, using the results of section 2 (eqns 
\req{GQh}-\req{AADh}). We will be interested only in the box diagrams
which contain \Gn-closed topological singular vectors built on \Gn-closed 
primaries; \ie\ of type $\kc^{(q)G}_{l,{\kp^G}}{\ }$, (see table 
\req{tabl2}), since only these types have a direct relation with the 
generic NS singular vectors via the topological (un)twistings. From 
the box diagrams one can deduce straightforwardly two different
mappings from the NS \svec s to the topological singular vectors,  
taking into account that in each box diagram the topological \svec\ of 
type $\kc^{(q)G}_{l,{\kp^G}}{\ }$ can be transformed into two NS \svec s
using the topological (un)twistings $T_W^{\pm }$ \req{twa}. 
These two NS \svec s are mirror-symmetric under
the exchange $H_m \to -H_m$ and $G^+_r \leftrightarrow G^-_r$; therefore
they have opposite U(1) charges and are located in mirror-symmetric
Verma modules:
\BE \kc^{(q)G}_{l,\,{\ket{\D,\,\htop}^G}} = 
 T_W^+ {\ } \ket{\chi_{NS}}^{(q)}_{l-q/2,\,{\ket{\D-\htop/2,\,\htop}}}   
 = T_W^- {\ }
 \ket{\chi_{NS}}^{(-q)}_{l-q/2,\,{\ket{\D-\htop/2,-\htop}}}\,,  
 \label{mtq} \EE

Let us start with the box diagrams which contain an uncharged singular 
vector of type $\kc^{(0)G}_{l,{\kp^G}}{\ }$ in the Verma module
$V(\ket{\D,\htop}^G)$. For non-zero conformal weight, $\D+l \neq 0$, the 
box diagram, shown in \req{diab0}, consists of 
singular vectors of the types $\kc^{(0)G}_{l,{\kp^G}}{\ }$ and
$\kc^{(-1)Q}_{l,{\kp^G}}{\ }$, at level $l$ in the Verma module 
$V(\ket{\D,\htop}^G)$, and singular 
vectors of the types $\kc^{(0)Q}_{l,{\kp^Q}}{\ }$ 
and $\kc^{(1)G}_{l,{\kp^Q}}{\ }$, also at level $l$ in the Verma module 
 $V(\ket{\D,-\htop-\ctop/3}^Q)$. 

\vskip .2in

\def\btggo  {\mbox{$\kc_{l,\, \ket{\D,\,\htop}^G}^{(0)G} $}}
\def\btqqo  {\mbox{$\kc_{l,\, \ket{\D,-\htop-{\ctop\over3}}^Q}^{(0)Q} $}}
\def\btqqm  {\mbox{$\kc_{l,\, \ket{\D,\,\htop}^G}^{(-1)Q} $}}
\def\btggp  {\mbox{$\kc_{l,\, \ket{\D,-\htop-{\ctop\over3}}^Q}^{(1)G} $}}

  \begin{equation}
  \begin{array}{rcl}
   \btggo &
  \stackrel{\Qz}{\mbox{------}\!\!\!\longrightarrow}
  & \btqqm \\[3 mm]
   \cA\,\updownarrow\ && \ \updownarrow\, \cA
  \\[3 mm]   \btqqo \! & \stackrel{\Gz}
  {\mbox{------}\!\!\!\longrightarrow} & \! \btggp
  \end{array} \label{diab0} \end{equation}

\vskip .2in
\noi
The arrows \Gn\ and \Qn\ can be reversed (up to constants) using \Qn\
and \Gn\ respectively; that is, the fermionic zero modes interpolate 
between two singular vectors, one charged and one uncharged, 
at the same level in the same Verma module. 

For $\D+l = 0$, the conformal weight of the singular 
vectors is zero, so that the corresponding 
arrows $\Qz$, $\Gz$ cannot be reversed, producing
{\it secondary} chiral singular vectors
$\kc_{l,\,\ket{-l,\,\htop}^G}^{(-1)G,Q}$ and
$\kc_{l,\,\ket{-l,-\htop-\ctop/3}^Q}^{(1)G,Q}$ on the right-hand side,
at level zero with respect to the singular vectors on the left-hand side.
The corresponding box diagram is therefore: 

\vskip .2in

\def\btggoc  {\mbox{$\kc_{l,\, \ket{-l,\,\htop}^G}^{(0)G} $}}
\def\btqqoc  {\mbox{$\kc_{l,\, \ket{-l,-\htop-{\ctop\over3}}^Q}^{(0)Q} $}}
\def\btqqmc  {\mbox{$\kc_{l,\, \ket{-l,\,\htop}^G}^{(-1)G,Q} $}}
\def\btggpc  {\mbox{$\kc_{l,\, \ket{-l,-\htop-{\ctop\over3}}^Q}^{(1)G,Q} $}}

  \begin{equation}
  \begin{array}{rcl}
   \btggoc &
  \stackrel{\Qz}{\mbox{------}\!\!\!\longrightarrow}
  & \btqqmc \\[3 mm]
   \cA\,\updownarrow\ && \ \updownarrow\, \cA
  \\[3 mm]   \btqqoc \! & \stackrel{\Gz}
  {\mbox{------}\!\!\!\longrightarrow} & \! \btggpc
  \end{array} \label{diab0ch} \end{equation}

\vskip .2in

The untwisting of the uncharged \svec\ 
$\kc^{(0)G}_{l,\,{\ket{\D,\,\htop}^G}}{\ }$, using $T_W^{\pm }$ \req{twa},
produces two uncharged mirror-symmetric NS \svec s 
located in mirror-symmetric Verma modules. Conversely, the twisting
of two uncharged mirror-symmetric NS \svec s, using $T_W^+$ and 
$T_W^-$ respectively, produces the same uncharged topological \svec\ of 
type $\kc^{(0)G}_{l,{\kp^G}}{\ }$. That is, one has the mappings: 
\BE \kc^{(0)G}_{l,\,{\ket{\D+\htop/2,\,\htop}^G}} = 
 T_W^+ {\ } \ket{\chi_{NS}}^{(0)}_{l,\,{\ket{\D,\,\htop}}}   
 = T_W^- {\ }
 \ket{\chi_{NS}}^{(0)}_{l,\,{\ket{\D,-\htop}}}\,,  \label{mtu} \EE

\noi
 where we have redefined $\D$ as the conformal weight of the NS primaries.
 The mappings from the NS \svec s to the remaining topological \svec s 
 in diagrams \req{diab0} and \req{diab0ch} can be derived now
 resulting in the following expressions:

\BE \begin{array}{lcl}

  {\ }\kc^{(0)Q}_{l,\,{\ket{\D+\htop/2,-\htop-{\ctop\over3}}^Q}} &=& 
\cA {\ } T_W^+ {\ } \ket{\chi_{NS}}^{(0)}_{l,\,{\ket{\D,\,\htop}}} \\  

  {\ }\kc^{(-1)Q}_{l,\,{\ket{\D+\htop/2,\,\htop}^G}} &=& 
\Qz {\ } T_W^+ {\ } \ket{\chi_{NS}}^{(0)}_{l,\,{\ket{\D,\,\htop}}} \\  

  {\ }\kc^{(1)G}_{l,\,{\ket{\D+\htop/2,-\htop-{\ctop\over3}}^Q}} &=& \cA 
 {\ }\Qz {\ } T_W^+ {\ } \ket{\chi_{NS}}^{(0)}_{l,\,{\ket{\D,\,\htop}}} \\  

  {\ }\kc^{(-1)G,Q}_{l,\,{\ket{-l,\,\htop}^G}} &=& \Qz {\ } 
T_W^+ {\ } \ket{\chi_{NS}}^{(0)}_{l,\,{\ket{-l-\htop/2,\,\htop}}} \\ 

  {\ }\kc^{(1)G,Q}_{l,\,{\ket{-l,-\htop-{\ctop\over3}}^Q}} &=& \cA {\ }\Qz 
  {\ } T_W^+ {\ } \ket{\chi_{NS}}^{(0)}_{l,\,{\ket{-l-\htop/2,\,\htop}}}   
 \label{cfu} \end{array} \EE

\vskip .15in
\noi
and similar expressions using 
$T_W^- \,\ket{\chi_{NS}}^{(0)}_{l,\,{\ket{\D,-\htop}}} $. Observe that
the two last mappings, to chiral \svec s, are not invertible since the 
arrows \Qn , \Gn\ in diagram \req{diab0ch} cannot be reversed. 

\vskip .15in

One finds similar box diagrams associated to the charge $q=1$ singular 
vector $\kc^{(1)G}_{l,{\kp^G}}{\ }$ in the Verma module
$V(\ket{\D,\htop}^G)$, as shown in \req{diab1} and
\req{diab1ch}. For $\D+l \neq 0$ the box diagram consists of \svec s 
of the types $\kc^{(1)G}_{l,{\kp^G}}{\ }$ and $\kc^{(0)Q}_{l,{\kp^G}}{\ }$
at the same level $l$ in the Verma module $V(\ket{\D,\htop}^G)$, and
\svec s of the types $\kc^{(-1)Q}_{l,{\kp^Q}}{\ }$ and 
$\kc^{(0)G}_{l,{\kp^Q}}{\ }$ also at level $l$ in the
Verma module $V(\ket{\D,-\htop-\ctop/3}^Q)$. 

\vskip .2in

\def\btggob  {\mbox{$\kc_{l,\, \ket{\D,\,\htop}^G}^{(1)G} $}}
\def\btqqob  {\mbox{$\kc_{l,\, \ket{\D,-\htop-{\ctop\over3}}^Q}^{(-1)Q} $}}
\def\btqqmb  {\mbox{$\kc_{l,\, \ket{\D,\,\htop}^G}^{(0)Q} $}}
\def\btggpb  {\mbox{$\kc_{l,\, \ket{\D,-\htop-{\ctop\over3}}^Q}^{(0)G} $}}

  \begin{equation} \begin{array}{rcl}
  \btggob &
  \stackrel{\Qz}{\mbox{------}\!\!\!\longrightarrow}
  & \btqqmb \\[3 mm]
   \cA\,\updownarrow\ && \ \updownarrow\, \cA
  \\[3 mm]  \btqqob \! & \stackrel{\Gz}
  {\mbox{------}\!\!\!\longrightarrow} & \! \btggpb
 \end{array} \label{diab1} \end{equation}
 
 \vskip .2in

For the case of zero conformal weight $\D+l=0$ the uncharged \svec s on
the right-hand side become secondary chiral singular vectors: 
 
\vskip .2in

\def\btggobc  {\mbox{$\kc_{l,\, \ket{-l,\,\htop}^G}^{(1)G} $}}
\def\btqqobc  {\mbox{$\kc_{l,\, \ket{-l,-\htop-{\ctop\over3}}^Q}^{(-1)Q}$}}
\def\btqqmbc  {\mbox{$\kc_{l,\, \ket{-l,\,\htop}^G}^{(0)G,Q} $}}
\def\btggpbc  {\mbox{$\kc_{l,\, \ket{-l,-\htop-{\ctop\over3}}^Q}^{(0)G,Q}$}}

  \begin{equation} \begin{array}{rcl}
  \btggobc &
  \stackrel{\Qz}{\mbox{------}\!\!\!\longrightarrow}
  & \btqqmbc \\[3 mm]
   \cA\,\updownarrow\ && \ \updownarrow\, \cA
  \\[3 mm]  \btqqobc \! & \stackrel{\Gz}
  {\mbox{------}\!\!\!\longrightarrow} & \! \btggpbc
 \end{array} \label{diab1ch} \end{equation}
 
 \vskip .2in

The untwisting of the charged \svec\ 
$\kc^{(1)G}_{l,\,{\ket{\D,\,\htop}^G}}{\ }$, using $T_W^{\pm }$ \req{twa},
produces two charged mirror-symmetric NS \svec s 
located in mirror-symmetric Verma modules. Conversely, the twisting
of two charged mirror-symmetric NS \svec s, using $T_W^+$ and $T_W^-$ 
respectively, produces the same charged $(|q|=1)$ topological \svec . 
For charge $q=1$ one finds the mappings:
\BE \kc^{(1)G}_{l,\,{\ket{\D+\htop/2,\,\htop}^G}} = 
 T_W^+ {\ } \ket{\chi_{NS}}^{(1)}_{l-1/2,\,{\ket{\D,\,\htop}}}   
 = T_W^- {\ }
 \ket{\chi_{NS}}^{(-1)}_{l-1/2,\,{\ket{\D,-\htop}}}\,. \label{mtch+} \EE

\noi
 where we have redefined $\D$ again, for convenience.
 The mappings from the NS \svec s to the remaining topological \svec s 
 in diagrams \req{diab1} and \req{diab1ch} are given by:

\vskip .15in

\BE \begin{array}{lcl}
\kc^{(-1)Q}_{l,\,{\ket{\D+\htop/2,-\htop-{\ctop\over3}}^Q}} &=& \cA {\ } 
T_W^+ {\ } \ket{\chi_{NS}}^{(1)}_{l-1/2,\,{\ket{\D,\,\htop}}} \\   

\kc^{(0)Q}_{l,\,{\ket{\D+\htop/2,\,\htop}^G}} &=& \Qz {\ } 
T_W^+ {\ } \ket{\chi_{NS}}^{(1)}_{l-1/2,\,{\ket{\D,\,\htop}}} \\  

\kc^{(0)G}_{l,\,{\ket{\D+\htop/2,-\htop-{\ctop\over3}}^Q}} &=& \cA {\ }
\Qz {\ } T_W^+ {\ } \ket{\chi_{NS}}^{(1)}_{l-1/2,\,{\ket{\D,\,\htop}}} \\  

\kc^{(0)G,Q}_{l,\,{\ket{-l,\,\htop}^G}} &=& \Qz {\ } 
T_W^+ {\ } \ket{\chi_{NS}}^{(1)}_{l-1/2,\,{\ket{-l-\htop/2,\,\htop}}} \\  

\kc^{(0)G,Q}_{l,\,{\ket{-l,-\htop-{\ctop\over3}}^Q}} &=& \cA {\ }\Qz {\ } 
T_W^+ {\ } \ket{\chi_{NS}}^{(1)}_{l-1/2,\,{\ket{-l-\htop/2,\,\htop}}}   
\label{cfch+} \end{array} \EE

\vskip .15in
\noi
and similar expressions using 
$T_W^- \,\ket{\chi_{NS}}^{(-1)}_{l-1/2,\,{\ket{\D,-\htop}}} $. As before,
the two last mappings, to chiral \svec s, are not invertible since the 
arrows \Qn , \Gn\ in diagram \req{diab1ch} cannot be reversed. 

\vskip .15in

Finally let us take a charge $q=-1$ singular vector of type
 $\kc^{(-1)G}_{l,{\kp^G}}{\ }$ in the Verma module
 $V(\ket{\D,\htop}^G)$. For $\D+l \neq 0$ the box diagram, shown 
 in \req{diab2}, consists of \svec s of the types
 $\kc^{(-1)G}_{l,{\kp^G}}{\ }$ and $\kc^{(-2)Q}_{l,{\kp^G}}{\ }$, at the
 same level $l$ in the Verma module $V(\ket{\D,\htop}^G)$, and 
 singular vectors of the types  
$\kc^{(1)Q}_{l,{\kp^Q}}{\ }$ and $\kc^{(2)G}_{l,{\kp^Q}}{\ }$ also at 
level $l$ in the Verma module $V(\ket{\D,-\htop-\ctop/3}^Q)$. 

\vskip .2in

\def\bbggob  {\mbox{$\kc_{l,\, \ket{\D,\,\htop}^G}^{(-1)G} $}}
\def\bbqqob 
 {\mbox{$\kc_{l,\, \ket{\D,-\htop-{\ctop\over3}}^Q}^{(1)Q} $}}
\def\bbqqmb 
 {\mbox{$\kc_{l,\, \ket{\D,\,\htop}^G}^{(-2)Q} $}}
\def\bbggpb 
 {\mbox{$\kc_{l,\, \ket{\D,-\htop-{\ctop\over3}}^Q}^{(2)G} $}}

  \begin{equation} \begin{array}{rcl}
  \bbggob &
  \stackrel{\Qz}{\mbox{------}\!\!\!\longrightarrow}
  & \bbqqmb \\[3 mm]
   \cA\,\updownarrow\ && \ \updownarrow\, \cA
  \\[3 mm]  \bbqqob \! & \stackrel{\Gz}
  {\mbox{------}\!\!\!\longrightarrow} & \! \bbggpb
 \end{array} \label{diab2} \end{equation}

\vskip .2in

As was pointed out before, the twisting 
of two charged mirror-symmetric NS \svec s, using $T_W^+$ and $T_W^-$ 
respectively, produces the same charged $(|q|=1)$ topological \svec . 
For charge $q=-1$ one finds the mappings:
\BE \kc^{(-1)G}_{l,\,{\ket{\D+\htop/2,\,\htop}^G}} = 
 T_W^+ {\ } \ket{\chi_{NS}}^{(-1)}_{l+1/2,\,{\ket{\D,\,\htop}}}   
 = T_W^- {\ }
 \ket{\chi_{NS}}^{(1)}_{l+1/2,\,{\ket{\D,-\htop}}}\,.\label{mtch-} \EE

 The mappings from the NS \svec s to the remaining topological \svec s 
 in diagram \req{diab2} result as follows:

 \BE \begin{array}{lcl} 
\kc^{(1)Q}_{l,\,{\ket{\D+\htop/2,-\htop-{\ctop\over3}}^Q}} &=& \cA {\ } 
T_W^+ {\ } \ket{\chi_{NS}}^{(-1)}_{l+1/2,\,{\ket{\D,\,\htop}}} \\  

\kc^{(-2)Q}_{l,\,{\ket{\D+\htop/2,\,\htop}^G}} &=& \Qz {\ } 
T_W^+ {\ } \ket{\chi_{NS}}^{(-1)}_{l+1/2,\,{\ket{\D,\,\htop}}} \\  

\kc^{(2)G}_{l,\,{\ket{\D+\htop/2,-\htop-{\ctop\over3}}^Q}} &=& \cA {\ }
  \Qz {\ } T_W^+ {\ } \ket{\chi_{NS}}^{(-1)}_{l+1/2,\,{\ket{\D,\,\htop}}}   
\label{cfch-} \end{array} \EE

\noi
and similar expressions using 
$T_W^- \,\ket{\chi_{NS}}^{(1)}_{l+1/2,\,{\ket{\D,-\htop}}} $.

\vskip .15in

Let us come back to diagram \req{diab2}.
For $\D+l=0$ the ``would be" secondary chiral singular vectors with $|q|=2$ 
simply do not exist, as follows from the results in tables \req{tabl2}
and \req{tabl3}. As a consequence, the singular vectors of types 
$\kc^{(-1)G}_{l,{\kp^G}}{\ }$ and $\kc^{(1)Q}_{l,{\kp^Q}}{\ }$ 
 ``become" actually chiral for zero conformal weight, \ie\ of types 
$\kc^{(-1)G,Q}_{l,{\kp^G}}{\ }$ and $\kc^{(1)G,Q}_{l,{\kp^Q}}{\ }$ 
instead, and the box diagram reduces to two chiral singular vectors,
connected by $\cA$. Thus one has the mappings: 

\BE \begin{array}{lclcl} \kc^{(-1)G,Q}_{l,\,{\ket{-l,\,\htop}^G}} &=& 
 T_W^+ {\ } \ket{\chi_{NS}}^{(-1)}_{l+1/2,\,{\ket{-l-\htop/2,\,\htop}}}   
 &=& T_W^- {\ }
\ket{\chi_{NS}}^{(1)}_{l+1/2,\,{\ket{-l-\htop/2,-\htop}}}\, \\
\kc^{(1)G,Q}_{l,\,{\ket{-l,-\htop-{\ctop\over3}}^Q}} &=& \cA {\ }
 T_W^+ {\ } \ket{\chi_{NS}}^{(-1)}_{l+1/2,\,{\ket{-l-\htop/2,\,\htop}}}   
 &=& \cA {\ } T_W^- {\ }
\ket{\chi_{NS}}^{(1)}_{l+1/2,\,{\ket{-l-\htop/2,-\htop}}}\,.\label{mtchc}
\end{array} \EE

An important observation here is the following.
The fact that $\kc^{(-1)G}_{l,\,{\ket{\D,\,\htop}^G}}{\ }$ becomes chiral
for $\D=-l$ implies necessarily that the NS singular vector
$\ket{\chi_{NS}}^{(-1)}_{l+1/2,\,{\ket{-l-\htop/2,\,\htop}}}$ is antichiral
(annihilated by $G^-_{-1/2}$), whereas the
NS \svec\ $\ket{\chi_{NS}}^{(1)}_{l+1/2,\,{\ket{-l-\htop/2,-\htop}}}$ 
is chiral (annihilated by $G^+_{-1/2}$). The reason is that 
$Q_0 = T_W^+\,G^-_{-1/2} = T_W^-\,G^+_{-1/2}$, so that the condition
of being annihilated by \Qn\ is transformed into the conditions of being
annihilated by $G^-_{-1/2}$ and $G^+_{-1/2}$, respectively, under
the (un)twistings $T_W^+$ and $T_W^-$. Thus we have found that charged NS 
\svec s $\ket{\chi_{NS}}^{(\pm 1)}_{l',\,{\ket{\D',\,\htop'}}}$ with
$\D'+l'= \pm {(\htop'\pm 1)\over2}$ are chiral (upper signs) and antichiral 
(lower signs). 

The uncharged chiral singular vectors are equivalent to charged chiral 
singular vectors, as was pointed out in section 2. Namely
\BE \kc^{(0)G,Q}_{l,\,{\ket{-l,\,\htop-1}^Q}} = 
\kc^{(-1)G,Q}_{l,\,{\ket{-l,\,\htop}^G}}\,, {\ } \qquad
\kc^{(0)G,Q}_{l,\,{\ket{-l,-\htop-{\ctop\over3}+1}^G}}=
\kc^{(1)G,Q}_{l,\,{\ket{-l,-\htop-{\ctop\over3}}^Q}}\,, \label{mtchu} \EE

\noi
by exchanging the primary states of the Verma module: 
$\ket{-l,\,\htop}^G = G_0 \, \ket{-l,\,\htop-1}^Q$ and
$\ket{-l,-\htop-{\ctop\over3}}^Q = Q_0 \, 
\ket{-l,-\htop-{\ctop\over3}+1}^G$.

Hence we have found invertible mappings from the NS \svec s to the 
chiral topological singular vectors. These are, in addition, simpler than 
the mappings \req{cfu} and \req{cfch+} deduced from diagrams 
\req{diab0ch} and \req{diab1ch}.
  
\vskip .2in

The mappings given by eqns. \req{mtu}, \req{cfu}, \req{mtch+}
\req{cfch+}, \req{mtch-}, \req{cfch-}, and \req{mtchc} 
turn into construction formulae for the topological
singular vectors just by expressing the NS singular vectors
$\ket{\chi_{NS}}^{(0)}$, $\ket{\chi_{NS}}^{(1)}$ and
$\ket{\chi_{NS}}^{(-1)}$
in terms of their corresponding construction formulae, given in
refs. \cite{Doerr1} and \cite{Doerr2}. 
The explicit expressions for the charged \svec s at level $k$ read 
\cite{Doerr1}
\BE \ket{\chi_{NS}}^{(\pm1)}_k= {\cal W}^{\pm} {\cal E}^{\pm}(k-1/2)
{\cal T}^{\pm}(k-1){\cal E}^{\pm}(k-3/2){\cal T}^{\pm}(k-2) ....
{\cal E}^{\pm}(1) {\cal T}^{\pm}(1/2) \Psi_0^{\pm} \,, \EE

\noi
where ${\cal E}^{\pm}(k)$ and ${\cal T}^{\pm}(k)$ 
are even and odd recursion step matrices, respectively, 
and ${\cal W}^{\pm}$ and $\Psi_0^{\pm}$ are
vectors, the latter depending on the initial low level \svec s. The
spectrum of $\D$ and $\htop$ for which the NS Verma modules 
$V_{NS}(\D,\,\htop)$ contain charged \svec s $\ket{\chi_{NS}}^{(\pm1)}_k$
is given (at least) by the zeroes of the NS determinant formula which are 
solutions to the vanishing planes $g_{\pm k}(\D,\,\htop)=0$ \cite{detNS}.

The explicit expressions for the uncharged \svec s at level 
$l={rs\over2}$ read \cite{Doerr2}
\BE \ket{\chi_{NS}}^{(0)}_{r,s}= 
\epsilon_{r,s}^{+}(t,\,\htop) {\ } \D_{r,s}(1,0){\ } + {\ }
\epsilon_{r,s}^{-}(t,\,\htop) {\ } \D_{r,s}(0,1)\,,
\EE

\noi
where $\D_{r,s}(1,0)$ and $\D_{r,s}(0,1)$ are two basis vectors taken from
the analytically continued Verma module,  
$t={3-\ctop\over3}$ parametrizes the central charge, and
\BE
\epsilon_{r,s}^{\pm}(t,\htop)= \prod_{m=1}^r
(\pm{s-rt\over2t}+{\htop\over t}\mp{1\over2}\pm m)\,\  ,\ r\in\oZ^+,
\,\, s\in2\oZ^+  \,.
\label{Dcond}
\EE

The spectrum of $\D$ and $\htop$ for which the NS Verma modules 
$V_{NS}(\D,\,\htop)$ contain uncharged \svec s 
$\ket{\chi_{NS}}^{(0)}_{r,s}$ is given (at least) by the zeroes of the NS 
determinant formula which are solutions to the quadratic vanishing 
surface $f_{r,s}(\D,\,\htop)=0$ \cite{detNS}.

The simultaneous vanishing of the two curves,
$\epsilon_{r,s}^+(t,\htop)=0$ and $\epsilon_{r,s}^-(t,\htop)=0$, 
leads to the appearance of two linearly independent uncharged NS 
singular vectors at the same level, in the same Verma module \cite{Doerr2}.
The topological twists $T_W^{\pm }$ \req{twa} let 
these conditions invariant, extending the existence of the two-dimensional 
space of singular vectors to the topological singular vectors of types
$\kc_{\kp^G}^{(0)G}$, $\kc_{\kp^G}^{(-1)Q}$, $\kc_{\kp^Q}^{(1)G}$ and
$\kc_{\kp^Q}^{(0)Q} \,$, as the generic uncharged NS \svec s are transformed 
necessarily into these four types of topological \svec s via the 
mappings \req{mtu} and \req{cfu}. The chiral types of singular vectors, 
which are related to particular, non-generic uncharged NS \svec s 
via the mappings \req{cfu}, do not admit two-dimensional 
spaces however\cite{DB2}. Nevertheless, they may appear as 
partners of \Gn-closed or \Qn-closed singular vectors in D\"orrzapf 
pairs since they are just particular cases of \Gn-closed 
and \Qn-closed singular vectors (see Appendix C in ref. \cite{BJI6}).

For the NS charged \svec s there are no two-dimensional spaces either
\cite{Doerr2}. This also implies the absence of two-dimensional
spaces for all types of topological \svec s connected to them generically 
via the mappings \req{mtch+}-\req{cfch-}; that is, for the eight types:
$\kc_{l,\ket{\phi}^G}^{(-2)Q}$, $\kc_{l,\ket{\phi}^Q}^{(-1)Q}$,
$\kc_{l,\ket{\phi}^G}^{(0)Q}$, $\kc_{l,\ket{\phi}^Q}^{(1)Q}$,
$\kc_{l,\ket{\phi}^G}^{(-1)G}$, $\kc_{l,\ket{\phi}^Q}^{(0)G}$,
$\kc_{l,\ket{\phi}^G}^{(1)G}$ and $\kc_{l,\ket{\phi}^Q}^{(2)G}$.

\vskip .17in
With the same reasoning, and taking into account that there are no
NS \svec s $\ket{\chi_{NS}}^{(q)}_{l}$ with $|q| \geq 2$ (as was
proved in ref. \cite{Doerr2}), one can prove the non-existence of
topological \svec s of types $\kc_{l,\ket{\phi}^G}^{(q)G}$ with
$|q| \geq 2$ (because of result \req{mtq}), and the non-existence 
of all types of topological \svec s in the ``would-be" box-diagrams
associated to $\kc_{l,\ket{\phi}^G}^{(q)G}$, $|q| \geq 2$; that is,
\svec s of the types $\kc_{l,\ket{\phi}^G}^{(q-1)Q}$,
$\kc_{l,\ket{\phi}^Q}^{(-q)Q}$, $\kc_{l,\ket{\phi}^Q}^{(1-q)G}$,
$\kc_{l,\ket{\phi}^Q}^{(1-q)G,Q}$ and $\kc_{l,\ket{\phi}^G}^{(q-1)G,Q}$,
with $|q| \geq 2$.

\section{Final Remarks}\lvm

We have written down one-to-one mappings between the NS \svec s and 16 types 
of topological singular vectors: the ones given in tables \req{tabl2} and 
\req{tabl3} with the exception of the no-label types. One can write many 
other (chains of) transformations among the topological \svec s \cite{BJI6},
apart from the ones given by the box diagrams, used in this work. 
However, for the purpose 
of mapping the NS \svec s to the topological \svec s, the additional
transformations are of little interest. In particular there are no 
mappings from the NS \svec s to the no-label topological \svec s nor
to the topological \svec s in no-label Verma modules. It seems that these
types of topological \svec s can only be mapped either to NS subsingular
vectors \cite{DB1} or to null non-highest weight descendants of NS
singular vectors.

Two important consequences of the analysis performed in this paper are:
i) the non-existence of topological \svec s of types
$\kc_{l,\ket{\phi}^G}^{(q)G}$ with $|q| \geq 2$,
and of all the ``would-be" topological \svec s in their box-diagrams,
and ii) the fact that the charged NS \svec s 
$\ket{\chi_{NS}}^{(\pm 1)}_{l,\,{\ket{\D,\,\htop}}}$ with
$\D+l= \pm {(\htop\pm 1)\over2}$
are chiral (upper signs) and antichiral (lower signs). 

\vskip .3in
\centerline{\bf Acknowledgements}

I am grateful to M. D\"orrzapf for reading the manuscript and for
several important suggestions.

\end{document}